\documentclass[%
 aip,
 amsmath,amssymb,
 reprint,%
]{revtex4-1}

\usepackage{graphicx}
\usepackage{dcolumn}
\usepackage{bm}

\usepackage[utf8]{inputenc}
\usepackage[T1]{fontenc}
\usepackage{mathptmx}
\usepackage{etoolbox}

\makeatletter
\def\@email#1#2{%
 \endgroup
 \patchcmd{\titleblock@produce}
  {\frontmatter@RRAPformat}
  {\frontmatter@RRAPformat{\produce@RRAP{*#1\href{mailto:#2}{#2}}}\frontmatter@RRAPformat}
  {}{}
}%
\makeatother
\begin{document}

\preprint{AIP/123-QED}

\title{Effect of interfacial damping on high-frequency surface wave resonance on a nanostrip-bonded substrate}
\author{Wenlou Yuan}
\author{Akira Nagakubo}
\author{Hirotsugu Ogi}
 \homepage{ogi@prec.eng.osaka-u.ac.jp\\
 Accepted for publication in \textit{Journal of Applied Physics} on December 10, 2021}
\affiliation{%
Graduate School of Engineering, Osaka University, 2-1 Yamadaoka, Suita, Osaka 565-0871, Japan 
}%

\date{\today}

\begin{abstract}
Since surface acoustic waves (SAW) are often generated on substrates to which nanostrips are periodically attached, it is very important to consider the effect of interface between the deposited strip and the substrate surface, which is an unavoidable issue in manufacturing. In this paper, we propose a theoretical model that takes into account the interface damping and calculate the dispersion relationships both for frequency and attenuation of SAW resonance. This results show that the interface damping has an insignificant effect on resonance frequency, but, interestingly, attenuation of the SAW can decrease significantly in the high frequency region as the interface damping increases. Using picosecond ultrasound spectroscopy, we confirm the validity of our theory; the experimental results show similar trends both for resonant frequency and attenuation in the SAW resonance. Furthermore, the resonant behavior of the SAW is simulated using the finite element method, and the intrinsic cause of interface damping on the vibrating system is discussed. These findings strongly indicate the necessity of considering interfacial damping in the design of SAW devices.

\end{abstract}

\maketitle

\section{Introduction}
Inspired by the invention of interdigital transducer (IDT) structure,\cite{1} a variety of SAW devices have been proposed, including sensors, \cite{2} filters,\cite{3} and oscillators \cite{4} all of which offer advantages such as small size, low energy consumption, and large dynamic range. To precisely design the SAW devices for various applications, it is important to study the propagation behaviors of SAWs generated by periodic structures. For example, Lin et al. used  picosecond ultrasound to measure the vibrations of nanostructures of gold stripes and dots patterned on glass.\cite{5} Maznev and Every analyzed the surface acoustic modes of a structure consisting of a periodic array of copper and SiO$_2$ wires on a silicon substrate. They analyzed the mass-loading effect and gave the physical explanation of a bandgap arising from the hybridization and avoided crossing of the Rayleigh and Sezawa modes.\cite{6,7} Although Maznev and Wright mentioned that SAW can propagate without radiating energy into the substrate in the specific zero-group-velocity modes,\cite{8} in the general situation, as shown in the research by Gelda \textit{et al.}, the mass loading of the grating on the substrate dominates dissipation by radiating energy from the surface into the bulk in the range of 1–100 GHz.\cite{9} 

Although the propagation characteristics of SAWs have been widely studied, the interface problem between strips and substrate has not been investigated deeply, because their values are highly ambiguous and very difficult to obtain both theoretically and experimentally. However, interface problems such as interface delamination and degradation of interfacial bond are very common issues in the fabrication of SAW devices, and it is very important to study their influences on SAW propagation. For example, Guillet \textit{et al.}\cite{10} studied the GHz dynamics of the elastic contact between a single metallic nanoparticle and a substrate and found that the frequency and the lifetime of this axial oscillation are related to the nanoparticle-substrate contact stiffness. Chang \textit{et al.}\cite{11} used ultrafast laser pulses to excite acoustic phonons in single gold nanodisks with variable titanium layer thicknesses and observed an increase in the phonon frequency in the nanodisk with a thicker adhesion layer because of stronger binding to the substrate. In our previous research, we also reveal that the bond strength at the interface, or the interface stiffness, has a significant influence on resonance frequency and attenuation of high-frequency SAW resonance. \cite{12} Although these works have shown the significance of considering interface problems, there is no systematic study for the effect of the interface damping on the SAW propagation behavior. 

Similar to the interface stiffness, the interface damping, which is caused by interface dislocations, anharmonic interface bonding, friction due to localized disbonding, and so on,\cite{13}  will cause significant impact on the SAW propagation behaviors. In this paper, we study the impact of the interface damping between nanostrips and substrate in high-frequency (up to  12 GHz) SAW resonances. By introducing complex interfacial stiffness, we theoretically analyze the surface wave modes and carry out numerical simulations. The results show that the interface damping insignificantly affects the frequency of the SAW resonance, but it can significantly reduce the SAW attenuation at high frequencies despite the presence of additional damping sources. Using picosecond ultrasound spectroscopy on permalloy strips attached silicon substrate, we experimentally  confirm the validity of our theory. Furthermore, we use the finite element method (FEM) for modal analysis to study the influence of interface damping in more detail. The results show that the interface damping has an obvious impact on attenuation of high-frequency SAWs, and it will be a key parameter in the design of SAW devices.

\section{Theory}
We propose a two-dimensional model with interface stiffness and interface damping as illustrated in Fig. 1, where the $x_1$ axis defines the in-plane direction, and the $x_3$ axis defines the out-of-plane direction. The strips are connected to the substrate with elastic springs and dampers. The spring constants per unit area along the in-plane and out-of-plane directions are denoted as $K_1$ and $K_3$, respectively. Also, the ratio of the dampers along the in-plane and out-of-plane directions are defined as $Q_1^{-1}$ and $Q_3^{-1}$, respectively. For simplicity, we consider the Si substrate to be an isotropic material, with a density of 2335 kg/m$^{3}$ and the aggregated elastic constants of $C_{11}$= 184.5 GPa and $C_{44}$ = 66.24 GPa. 

Although Yi \textit{et al.} \cite{14} indicated significant contribution of damping in the bulk to the mechanical relaxation for a single Au nanostructure (a lossy material), the interface area between the nanostructure and substrate is relatively small. For periodic nanostrips on substrate studied here, the total interface area becomes larger, and damping inside the nanostrips will not be significant because the SAW wavelength is much larger than the nanostrip thickness; the contribution of the strain energy inside the nanostrip to the total strain energy of SAW will be insignificant. Since the purpose of this paper is to study how the interface damping affects the radiation of energy from the surface to the bulk, we have neglected the effect of the intrinsic damping in the nanostrips for simplicity.

\begin{figure}[tp]
\begin{center}
\includegraphics[width=75mm]{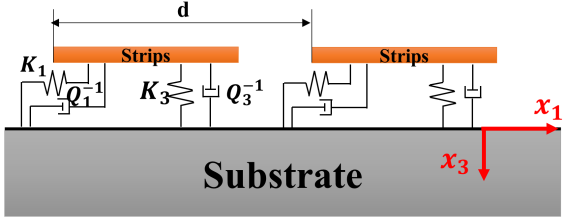}
\end{center}
\caption{Two-dimensional model with interface stiffness and interface damping.}
\label{1}
\end{figure}

Based on our previous model with the interface stiffness, we can find the wavenumber-frequency relationship with interface damping in the similar way.

First, displacements in the substate $u_1$ and $u_3$ can be expressed by superimposing the partial waves with the in-plane Bloch-harmonic wavenumbers $k_1^n$ ($n$ = 0, $\pm$1, $\pm$2,…)

\begin{equation}
\begin{aligned}
&u_{1}=\sum_{n}\left(A_{n} k_{1}^{n} e^{i\left(k_{1}^{n} x_{1}+k_{3}^{L n} x_{3}-\omega t\right)}+B_{n} k_{3}^{S n} e^{i\left(k_{1}^{n} x_{1}+k_{3}^{S n} x_{3}-\omega t\right)}\right) \\
&u_{3}=\sum_{n}\left(A_{n} k_{3}^{L n} e^{i\left(k_{1}^{n} x_{1}+k_{3}^{L n} x_{3}-\omega t\right)}-B_{n} k_{1}^{n} e^{i\left(k_{1}^{n} x_{1}+k_{3}^{S n} x_{3}-\omega t\right)}\right)
\end{aligned}
\end{equation}
Here,
\begin{equation}
k_{1}^{n}=k_{1}+\frac{2 \pi n}{d}
\end{equation}
$k_{3}^{L n}$ and $k_{3}^{S n}$ are wavenumbers along the $x_3$ direction for longitudinal and shear waves, respectively, and they are given by:

\begin{equation}
k_{3}^{L, S_{n}}= \begin{cases}i \sqrt{\left(k_{1}^{n}\right)^{2}-\frac{\omega^{2}}{\left(C_{L, S}\right)^{2}}}, & \left(k_{1}^{n}\right)^{2}>\frac{\omega^{2}}{\left(C_{L, S}\right)^{2}}, \\ i \sqrt{\frac{\omega^{2}}{\left(C_{L, S}\right)^{2}}-\left(k_{1}^{n}\right)^{2}}, & \left(k_{1}^{n}\right)^{2}<\frac{\omega^{2}}{\left(C_{L, S}\right)^{2}}.\end{cases}
\end{equation}
$C_{L, S}$ denotes longitudinal-wave ($C_L$) or shear wave ($C_S$) velocity in the substrate. The displacements of the strip can also be expanded with Bloch-harmonic terms:

\begin{equation}
\begin{aligned}
&\tilde{u}_{1}=\sum_{n}\left(C_{n} e^{i\left(k_{1}^{n} x_{1}-\omega t\right)}\right), \\
&\tilde{u}_{3}=\sum_{n}\left(D_{n} e^{i\left(k_{1}^{n} x_{1}-\omega t\right)}\right).
\end{aligned}
\end{equation}
The boundary conditions of the balance between the stress and spring force on the substrate surface take the forms:
\begin{equation}
\begin{gathered}
\rho C_{S}^{2}\left(\frac{\partial u_{1}}{\partial x_{3}}+\frac{\partial u_{3}}{\partial x_{1}}\right)-\widetilde{K}_{1}\left(u_{1}-\tilde{u}_{1}\right)=0, \\
\rho\left(C_{L}^{2}-2 C_{S}^{2}\right) \frac{\partial u_{1}}{\partial x_{1}}+\rho C_{L}^{2} \frac{\partial u_{3}}{\partial x_{3}}-\widetilde{K}_{3}\left(u_{3}-\tilde{u}_{3}\right)=0,
\end{gathered}
\end{equation}
where $\rho$ is the mass density of the substrate, and $\widetilde{K}_{1}$ and $\widetilde{K}_{3}$ are the complex elastic stiffnesses. Here, we interpret the interface damping as the ratio of imaginary to real parts of the complex elastic stiffness:\cite{15}     

\begin{equation}
\widetilde{K}_{j}=K_{j}\left(1+i Q_{j}^{-1}\right), \quad j=1 \textrm{ or }3.
\end{equation}
Here, ${K}_{j}$ is the interface stiffness and $Q_{j}^{-1}$ represents the interface damping. Another boundary condition is the balance between the inertia force and the spring force at the strip:

\begin{equation}
\rho_{s}\left(x_{1}\right) \frac{\partial^{2} \widetilde{u}_{j}}{\partial t^{2}}-\widetilde{K}_{j}\left(u_{j}-\tilde{u}_{j}\right)=0, \quad j=1 \textrm{ or }3.
\end{equation}
Here $\rho_{s}\left(x_{1}\right)$ represents the area mass density for the attached strips, and it can be expanded with the Bloch-harmonic components as:

\begin{equation}
\rho_{s}\left(x_{1}\right)=\sum_{n=-\infty}^{\infty} \rho_{s}^{n} \operatorname{exp}\left(i \frac{2 \pi n}{d} x_{1}\right).
\end{equation}
Where,

\begin{equation}
\rho_{s}^{n}=\rho h \times \begin{cases}\frac{1}{\pi n} \sin \left(\frac{\pi n w}{d}\right), & n \neq 0 \\ \frac{w}{d}, & n=0\end{cases}
\end{equation}
$h$ and $w$ are height and width of the strip, respectively. Substituting the displacements and mass distribution into the boundary conditions yields the linear equations for individual $n$ values:

\begin{equation}
\begin{aligned}
&\left(2 i k_{1}^{n} k_{3}^{L n}-\frac{\widetilde{K}_{1} k_{1}^{n}}{\rho C_{S}^{2}}\right) A_{n}+\left[i\left\{\left(k_{3}^{S n}\right)^{2}-\left(k_{1}^{n}\right)^{2}\right\}-\frac{\widetilde{K}_{1} k_{3}^{S n}}{\rho C_{S}^{2}}\right] B_{n} \\
&+\left(\frac{\widetilde{K}_{1}}{\rho C_{S}^{2}}\right) C_{n}=0, \\
&\left(i\left\{\frac{\omega^{2}}{C_{S}^{2}}-2\left(k_{1}^{n}\right)^{2}\right\}-\frac{\widetilde{K}_{3} k_{3}^{L n}}{\rho C_{S}^{2}}\right) A_{n}+\left[-2 i k_{1}^{n} k_{3}^{L n}+\frac{\widetilde{K}_{3} k_{1}^{n}}{\rho C_{S}^{2}}\right] B_{n} \\
&+\left(\frac{\widetilde{K}_{3}}{\rho C_{S}^{2}}\right) D_{n}=0,  \\
&\widetilde{K}_{1} k_{1}^{n} A_{n}+\widetilde{K}_{1} k_{3}^{S n} B_{n}-\widetilde{K}_{1} C_{n}+\omega^{2} \sum_{m=-\infty}^{\infty} \rho_{s}^{n-m} C_{m}=0, \\
&\widetilde{K}_{3} k_{3}^{L n} A_{n}-\widetilde{K}_{3} k_{1}^{n} B_{n}-\widetilde{K}_{3} D_{n}+\omega^{2} \sum_{m=-\infty}^{\infty} \rho_{s}^{n-m} D_{m}=0.
\end{aligned}
\end{equation}
To obtain the meaningful solutions for non-zero coefficients of $A_n$, $B_n$, $C_n$ and $D_n$, the determinant of the matrix formed by the above system equations should be zero. Thus, it provides relationships among wavenumber $k_1$, frequency $\omega$, and parameters of the structure. For the harmonics in the equation, we used terms up to |n|,|m| = 10, which are sufficient orders for accurate calculations. \cite{12} 

In the following numerical simulation, the above function is solved by seeking the frequency $\omega$ and the imaginary part of the surface-wavenumber Im$\left(k_1\right)$, which corresponds to the attenuation of leaky surface waves, for a given real part of the surface-wave-number Re$\left(k_1\right)$. Then, we obtain the relationship between the reciprocal period and the SAW resonance frequency as well as that between the reciprocal period and attenuation. 

\section{Experiment}
Nanostrip structures were fabricated on a $\left(001\right)$ Si substrate by the electron-beam lithography. A thin titanium layer (1 nm) was first deposited on the substrate to improve the bond strength between nanostrips and substrate, and then, permalloy (Fe$_{20}$Ni$_{80}$, $\rho$ = 8693 kg/m$^{3}$) layer was deposited with a thickness of about 20 nm. A thin silica layer (2 nm) was also deposited on the top of the permalloy layer to prevent oxidation. After lithographic procedures, five groups of nanostrip structures with different widths between 150 nm and 350 nm were fabricated. Since the width of gaps is the same as the width of the nanostrip, the period of the nanostrip structures is between 300 nm and 700 nm, as shown in Fig. 2.

The resonance of the SAW was generated and detected by the picosecond ultrasound spectroscopy. We used a mechanical delay line and moved the stage so as to obtain the reflectivity with 1.067 ps step. Because our mechanical delay-line system is highly optimized and it causes very little disturbance to the optical lines during the stage movement, our measurement system hardly affects the attenuation measurement. Details of the optics can be found in our previous works. \cite{12,16,17}   

\begin{figure}[tp]
\begin{center}
\includegraphics[width=75mm]{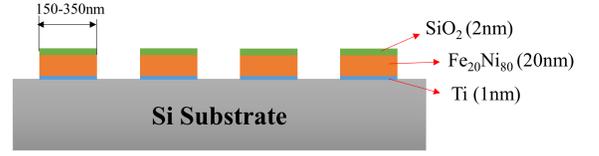}
\end{center}
\caption{Schematic of nanostrip-attached substrate structure used in experiment.}
\label{1}
\end{figure}

Fig. 3(a) shows the raw experiment data of the time-reflectivity change of the probe light for the structure with the strip period $d$ = 700 nm, and Figure 3(b) shows its Fourier spectrum, showing the SAW resonance at 6.4 GHz. Then, we used the bandpass filter designed by MATLAB Filter Design Toolbox (pass band frequency:5-15GHz) to suppress the noise signal. We used the same wide-band filter for all of the experiment signals to minimize the filtering effect on the attenuation determination.  Also, to evaluate the attenuation coefficient $\alpha_{t}$ from the decay oscillation, the damping sinusoidal function:

\begin{figure}[tp]
\begin{center}
\includegraphics[width=90mm]{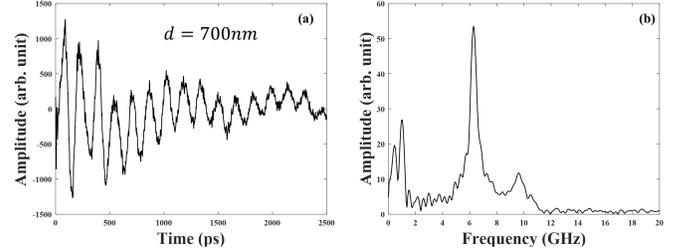}
\end{center}
\caption{(a) The raw experiment data of the reflectivity change for the specimen with $d$ = 700 nm, and (b) corresponding Fourier spectrum.}
\label{1}
\end{figure}

\begin{equation}
A=A_{0} \exp \left(-\alpha_{t} t\right) \sin (\omega t+\varphi)
\end{equation}
was fitted to the measurement. The R-square value of the curve fitting was higher than 0.90.

\section{Results}
To show the impact of the interface damping clearly, the numerical simulation without the interface damping is firstly performed for comparison.
The in-plane and out-of-plane stiffness values of the strip are simply estimated by dividing the shear modulus $K_1^M$ and Young’s modulus $K_3^M$ of the strip by the strip height, respectively, using the elastic constants of permalloy $C_{11}$ = 265 GPa and $C_{44}$ = 66 GPa.\cite{18} Since the in-plane and out-of-plane interface stiffnesses $K_1$ and $K_3$, respectively, will be lower than $K_1^M$ and $K_3^M$, we introduce the parameter $p$ (smaller than 1), which represents the bond strength at the interface. The relationship between interface stiffness and interface bond strength is then given by:
\begin{equation}
K_{1}=p K_{1}^{M} \quad K_{3}=p K_{3}^{M}
\end{equation}
Here, we assume that $p$ is the same in different directions for simplicity, because this assumption will not change the principal conclusion. Since the interface bond strength is highly ambiguous, we use different $p$ values in the  numerical calculation to investigate its influence. 

Figure 4 shows the result of numerical simulations of SAW resonance frequency and attenuation with different interface bond strengths as well as the experiment results (solid squares). The broken line denotes the Rayleigh-wave resonance ($v$ = 4870 m/s).
\begin{figure}[tp]
\begin{center}
\includegraphics[width=75mm]{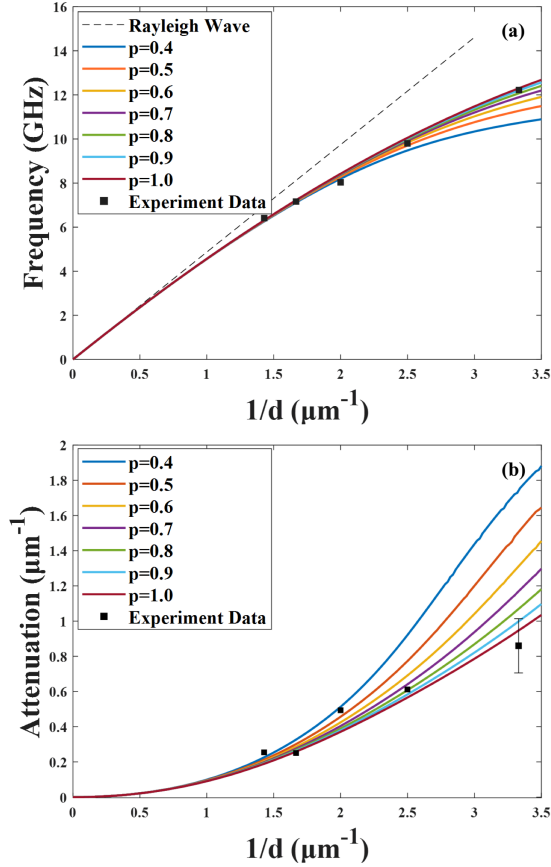}
\end{center}
\caption{Relationship of the reciprocal period with (a) the resonance frequency and (b) that with the attenuation coefficient with different interface bond strengths.}
\label{1}
\end{figure}
Figure 4(a) shows that the resonant frequency of SAW decreases with decreasing the adhesive strength, which is consistent with the conclusion of our previous study. Also, the bond strength that best agrees with the experimental results is $p$ = 0.8, which gives the interface stiffnesses of $K_{1}=2.64 \times 10^{18} \mathrm{~J} / \mathrm{m}^{4}$ and $K_{3}=7.04 \times 10^{18} \mathrm{~J} / \mathrm{m}^{4}$, which will be used in the following simulation with the interface damping.

The attenuation coefficient for distance $\alpha_{x}$ is calculated through $\alpha_{x}=\alpha_{t} / v_{s}$, where $\alpha_{t}$ denotes the measured attenuation coefficient for time, and $v_s$ is the group velocity of SAW, which is obtained through the dispersion relationship (Fig. 4(a)).\cite{19} Fig. 4(b) compares the calculations without the interface damping with experiments.  Focusing on the most suitable interface-bond strength of $p$ = 0.8, the experiments for attenuation are larger than the calculations in the low frequency region.  This is expected because we only consider the attenuation caused by bulk radiation as the energy dissipation mechanism of SAW, and other dissipation mechanisms, including the phonon-phonon interaction (Akhiezer mechanism),\cite{20} thermoelastic damping,\cite{21} phonon-electron interaction \cite{22} etc. could also contribute to the SAW attenuation to some extent. Based on the previous research, \cite{9} the attenuation caused by these mechanisms will contribute at least 10\%  to the whole energy dissipation near 10 GHz. However, in the high-frequency region beyond 10 GHz, the attenuation in the experiment is significantly lower than the calculation, which cannot be explained by the existing theory. 

We then introduce the interface damping into the numerical simulation, and study its effects on the SAW resonance behavior. Since the interface damping in different directions will have different effects, we independently investigate these effects. Figure 5 shows the effect of $Q_1^{-1}$ value. The red dashed line is the  simulation result without the interface damping and with bond strength of $p$ = 0.8 (previous theory). The colored solid lines are the simulation results of the proposed model with different in-plane interface-damping values. It is shown that SAW resonance frequency remains nearly unchanged even with the presence of the interface damping (Fig. 5(a)), which is expected, because in a vibration system, normally, damping should not influence the resonance frequency significantly. However, interestingly, as shown in Fig. 5(b), the SAW attenuation can be significantly reduced in the high frequency region due to the interface damping, which is in agreement with the experiment. The similar phenomenon is also found in the simulations with different $Q_3^{-1}$  values as shown in Fig. 6, although the in-plane interface damping $Q_1^{-1}$ has a larger impact on the SAW attenuation.

\begin{figure}[tp]
\begin{center}
\includegraphics[width=75mm]{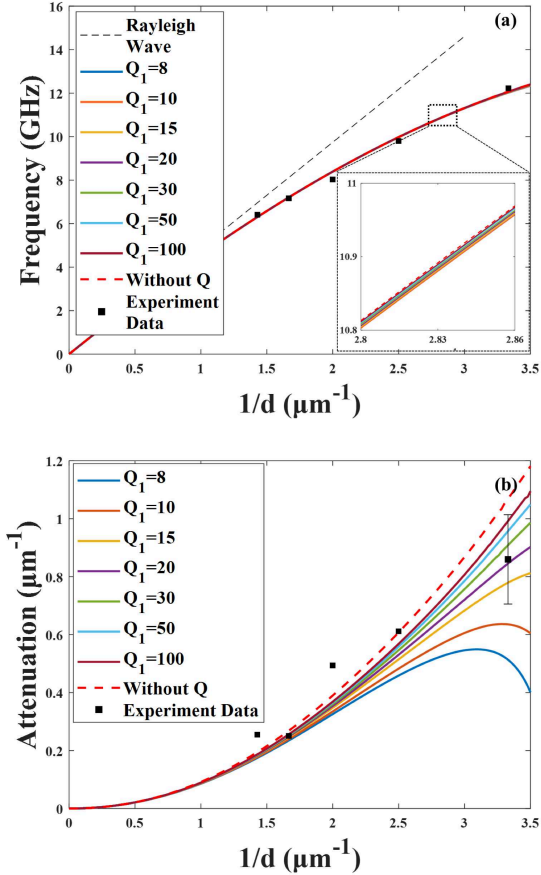}
\caption{Relationship of the reciprocal period with (a) the resonance frequency and (b) that with the attenuation coefficient with different in-plane interface damping $Q_1$ when the out-of-plane interface damping $Q_3$ is fixed to 10.}
\label{1}
\end{center}
\end{figure}

\begin{figure}
\begin{center}
\includegraphics[width=75mm]{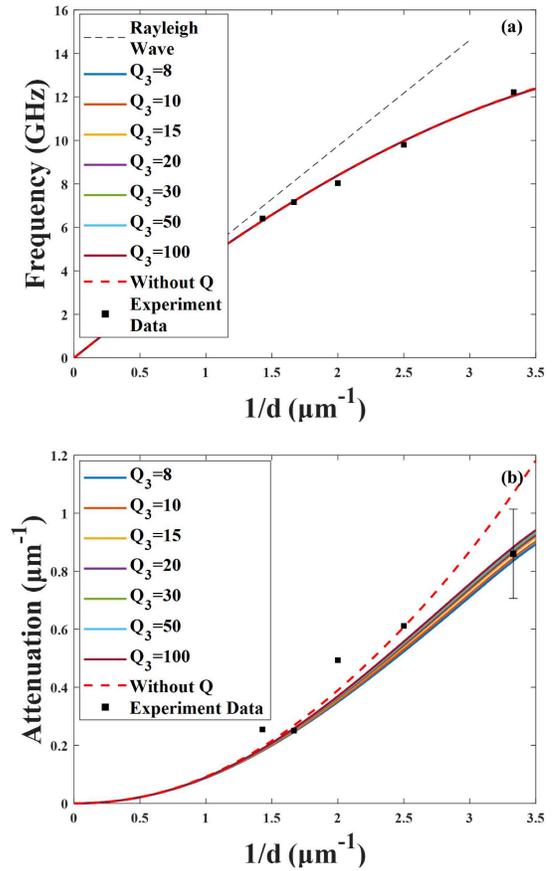}
\caption{Relationship of the reciprocal period with (a) the resonance frequency and (b) that with the attenuation coefficient with different out-of-plane interface damping $Q_3$ when the in-plane interface damping $Q_1$ is fixed to 20.}
\label{1}
\end{center}
\end{figure}

This is very interesting because with the addition of a damping mechanism we would expect the attenuation of the system to increase. However, the interface damping seems to suppress the leakage of the SAW energy into the bulk, and this mechanism has never been indicated. Furthermore, the simulation results based on our new model favorably explain the experiments with $Q_1$ = 20 and $Q_3$ = 15, indicating that the interface damping should be considered in discussing the SAW attenuation at high frequencies. 

\section{Finite Element Analysis}
In order to further explore the effect of the interface damping and also to validate our theory, we perform a modal analysis using FEM. \cite{23} The model to be analyzed is a two-dimensional one with permalloy strip of period $d$  attached on the silicon substrate of $10d$ thick. The material parameters are the same as in the above numerical calculation. 

For the restrictions, the bottom of the substrate is fixed because the thickness of the substrate is much larger than the wavelength as well as grating dimensions. The periodic Bloch condition is set on the left and right sides of the single cell to modify the entire nanostructure.\cite{24} The mesh dimensions were varied from 0.06045 nm near the top of the cell containing the strip and substrate to 70 nm in the bulk region of the cell. 

The results of the FEM analysis with the strips tightly attached to the substrate are shown in Fig. 7. In the FEM analysis, two principal modes are found for the SAW resonance.  One is the symmetric mode as shown in Fig. 7(a), and the other is the asymmetric mode as shown in Fig. 7(b). Because the SAW resonance is excited through thermal expansion due to irradiation with the pump light, it is expected that only symmetric modes like Figs. 7(b) and (c) are excited.  The resonance frequencies of SAW for $d$ = 300 and 500 nm can be then determined to be 14.243 and 9.1874 GHz, respectively.

 \begin{figure}[tp]
\begin{center}
\includegraphics[width=70mm]{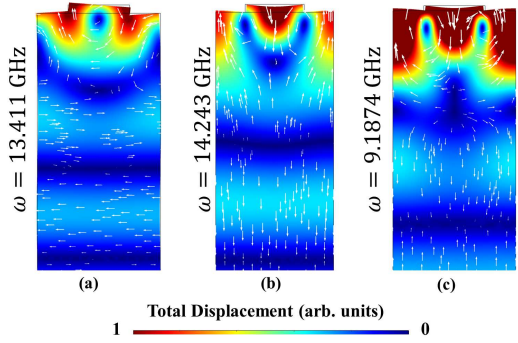}
\end{center}
\caption{Total-displacement images calculated by the FEM simulation for the model with strips tightly attached on the substrate.  (a) Asymmetric mode with period $d$ = 300 nm, (b) symmetric mode with period $d$ = 300 nm, and (c) symmetric mode with period $d$ = 500 nm.}
\label{1}
\end{figure}

For the FEM simulation with the interface stiffness and interface damping, a thin elastic layer is introduced to the interface between the strips and substrate, and we can set the elastic constant and the damping factor to this thin elastic layer to imitate the interface stiffness and interface damping.

\begin{figure}[tp]
\begin{center}
\includegraphics[width=75mm]{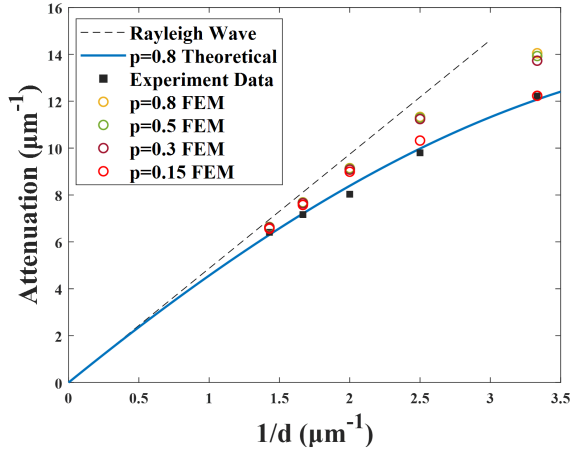}
\end{center}
\caption{Comparison of the FEM calculation and the theoretical calculation with different interface stiffnesses.}
\label{1}
\end{figure}

The FEM simulation result with different interface stiffness without interface damping is calculated for comparison first, as shown in Fig. 8, where the solid line is the theoretical calculation result when $p$ = 0.8 and the color circles are the FEM simulation results with different interface stiffnesses. It is clearly shown that reduction of the interface stiffness decreases the SAW resonance frequency, which agrees with our theoretical model. However, the SAW resonance frequency calculated by FEM is higher than that calculated by the theoretical method. This is acceptable
because in the theoretical calculation, only the mass-loading effect of the strips is involved, whereas, in the FEM simulation, the elasticity of strips is also involved, which will contribute to the increasing in the SAW resonance frequency. \cite{25}
Here, for $p$ = 0.15, the FEM simulation results  are in the best agreement with the experimental results, and this value will be used in the following FEM simulation involving the interface damping.
\begin{figure}[tp]
\begin{center}
\includegraphics[width=65mm]{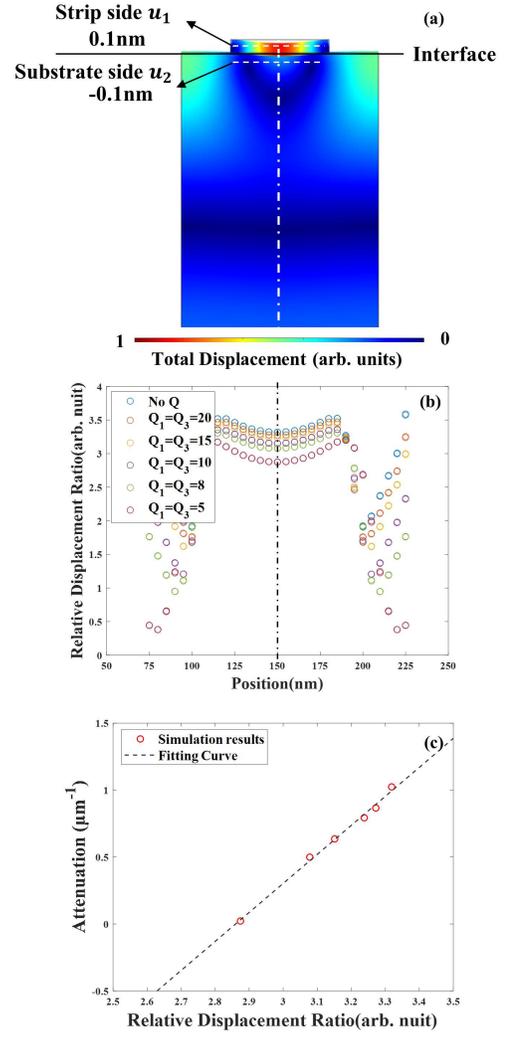}
\end{center}
\caption{(a) FEM simulation result when $d$ = 300 nm, $p$ = 0.15, and $Q_1$ = $Q_3$ = 5.  The white dashed lines indicate the locations where the total displacements are examined.  (b) The relative displacement ratio at the interface with different interface damping.  (c) Relationship between SAW attenuation calculated by theoretical model and the relative displacement ratio calculated by FEM.}
\label{1}
\end{figure}

Both the interface stiffness and interface damping are then introduced to the FEM simulation. In the previous study, it is shown that attenuation of SAW has a close relationship with the displacement distributions at the boundary between the IDT region and substrate. When they are very close, SAW is generated efficiently and less energy dissipates into the bulk, so that the attenuation of SAW is suppressed. \cite{26} Similarly, it is important to study the relative displacement between the strips and substrate in the SAW resonance mode, which will be helpful to illustrate the attenuation behavior at the SAW resonance. Although we cannot obtain the exact displacements in the strip and substrate in the modal analysis, it is useful to calculate their ratio to illustrate the essence. The result is shown in Fig. 9.

We calculated the total-displacement ratio $u_{1} / u_{2}$ between the displacement on strips side $u_{1}$ and substrate side $u_{2}$ at close to the interface (0.1nm), as shown in Fig. 9(a). The results are shown in Fig. 9(b), where colored circles denote the FEM simulation results for the total-displacement ratio at the interface between the strip and substrate. With increasing the interface damping, the displacement ratio between strips and substrate decreases in most parts of the interface. Furthermore, the relative displacement ratio at the center part of the interface (the chain line in Fig. 9(b)) matches well with the attenuation calculated in theoretical model (Fig. 9(c)), which will explain the reduction of the SAW attenuation: the interface damping reduces the relative displacement between strips and substrate, which will reduce the traction force and contribute to the improvement of the SAW generation efficiency and the reduction of bulk radiation.

\section{Conclusion}
In this paper, we study the impact of the interface damping between nanostrips and substrate in high-frequency SAW resonances. The numerical simulation results show that the interface damping will lead to the reduction of SAW attenuation significantly at high frequencies. Also, using picosecond ultrasound spectroscopy, we confirm the validity of our theory; the experimental results show the similar trends both for resonant frequency and attenuation of SAW. Then, by using the FEM simulation, it is shown that with the increase of interface damping, the relative displacement between the strips and substate decreases, which contributes to the reduction of the SAW attenuation. This work shows the importance of considering interface damping in SAW propagation researches and in designing a high frequency SAW device. We will extend our model to include the intrinsic material damping and establish a comprehensive model for analysis in our future study.

\section*{Acknowledgments}
This study was supported by JSPS KAKENHI Grant No. JP19H00862 and JST SPRING Grant No. JPMJSP2138. 

\section*{DATA AVAILABILITY}
The data that support the findings of this study are available from the corresponding author upon reasonable request.


\begin{thebibliography}{99}
\bibitem{1} R. M. White and F. W. Voltmer, Direct piezoelectric coupling to surface elastic waves. Appl. Phys. Lett. \textbf{7}, 314 (1965).
\bibitem{2} W. P. Jakubik, M. W. Urbańczyk, S. Kochowski and J. Bodzenta, Sens. Actuat. B: Chem. \textbf{82}, 265 (2002).
\bibitem{3} T. Nishihara, H. Uchishiba, O. Ikata and Y. Satoh, Jap. J. Appl. Phys.  \textbf{34}, 2688 (1995).
\bibitem{4} T. E. Parker and G. K. Montress, IEEE Trans. Ultrason. Feeroelectr. Freq. Control. \textbf{35}, 342 (1988).
\bibitem{5} H. N. Lin, H. J. Maris, L. B. Freund, K. Y. Lee, H. Luhn and D. P. Kern, J. Appl. Phys. \textbf{73}, 37 (1993).
\bibitem{6} A. A. Maznev, Phys. Rev. B \textbf{78}, 155323 (2008).
\bibitem{7} A. A. Maznev and A. G. Every, J. Appl. Phys.\textbf{106}, 113531 (2009).
\bibitem{8} A. A. Maznev and O. B. Wright, J. Appl. Phys.\textbf{105}, 123530 (2009).
\bibitem{9} D. Gelda, J. Sadhu, M. G. Ghossoub, E. Ertekin and S. Sinha, J. Appl. Phys. \textbf{119}, 164301 (2016).
\bibitem{10} Y. Guillet and B. Audoin, Phys. Rev. B \textbf{86}, 035456 (2012).
\bibitem{11} W. Chang, F. Wen, D. Chakraborty, M. Su, Y. Zhang, B. Shuang, P. Nordlander, J. E. Sader, N. J. Halas and S. Link, Nat. Commun.\textbf{6}, 1 (2015).
\bibitem{12} H. Ogi, S. Masuda, A. Nagakubo, N. Nakamura, M. Hirao, K. Kondou and T. Ono, Phys. Rev. B \textbf{93}, 024112 (2016).
\bibitem{13} J. P. Hirth, J. Phys. Chem. Soli. \textbf{55}, 985 (1994).
\bibitem{14} C. Yi, M. Su, P. D. Dongare, D. Chakraborty, Y. Cai, D. M. Marolf, R. N. Kress, B. Ostovar, L. J. Tauzin, F. Wen, W. Chang, M. R. Jones, J. E. Sader,N. J. Halas and Stephan Link, Nano Lett. \textbf{18}, 3494 (2018).
\bibitem{15} H. Ogi, T. Ohmori, N. Nakamura and M. Hirao, J. Appl. Phys \textbf{100}, 053511 (2006).
\bibitem{16} H. Ogi, A. Yamamoto, K. Kondou, K. Nakano, K. Morita, N. Nakamura, T. Ono, and M. Hirao, Phys. Rev. B  \textbf{82}, 155436 (2010).
\bibitem{17} A. Nagakugo, K. Adachi, T. Nishihara, and H. Ogi, Appl. Phys. Express \textbf{13}, 016504 (2020)
\bibitem{18} C. Rossignol, B. Perrin, B. Bonello, P. Djemia, P. Moch and H. Hurdequint, Phys. Rev. B  \textbf{70}, 094102 (2004).
\bibitem{19} M. E. Siemens, Q. Li, M. M. Murnane, H. C. Kapteyn, R. Yang, E. H. Anderson and K. A. Nelson, Appl. Phys. Lett. \textbf{94}, 093103 (2009).
\bibitem{20} B. Daly, K. Kang, Y. Wang, and D. G. Cahill, Phys. Rev. B \textbf{80}, 174112 (2009).
\bibitem{21} H. Maris, Phys. Rev. \textbf{188}, 1308 (1969).
\bibitem{22} V. J. Gokhale and M. Rais-Zadeh,  Sci. Rep. \textbf{4}, 1 (2014).
\bibitem{23} COMSOL Multiphysics finite element software, Version 5.6, COMSOL AB, Sweden.
\bibitem{24} D. Nardi, F. Banfi, C. Giannetti, B. Revaz, G. Ferrini, and F.Parmigiani, Phys. Rev. B \textbf{80}, 104119 (2009).
\bibitem{25} R. M. Taziev, IEEE Trans. Ultrason. Feeroelectr. Freq. Control.\textbf{54}, 2060 (2007).
\bibitem{26} S. Kaiko, K. Hishinuma and Y. Nakagawa IEEE Trans. J. Appl. Phys \textbf{87}, 1440 (2000).

\end{thebibliography}
\end{document}